\documentstyle[NATO,numreferences,epsfig]{crckapb}
\begin{opening}
\title{High Temperature Superconducting Radio Frequency Coils for NMR Spectroscopy
and Magnetic Resonance Imaging }
\author{Steven M. Anlage}

\institute{Center for Superconductivity Research, and 
Materials Research Science and
Engineering Center, 
Department of Physics, 
University of Maryland, 
\mbox{College Park, Maryland 20742-4111 USA}}
\end{opening}

\begin{document}

\begin{abstract}
High-temperature superconductivity has a significant opportunity to improve
the performance of nuclear magnetic resonance (NMR) spectroscopy and
magnetic resonance imaging (MRI) systems. The low rf losses and low operating
temperatures 
of superconducting coils allow them to improve the signal-to-noise ratio in
applications where the system noise dominates that of the sample under
study. These improvements translate into new capabilities and shorter
measurement times for NMR and MRI systems. In this pedagogical paper, we
discuss the implementation and impact of high-T$_c$ superconducting thin
films in rf coils for NMR and MRI applications.
\end{abstract}

\section{Introduction}

Nuclear Magnetic Resonance (NMR) has been employed for more than 50 years to
study the physics and chemistry of atoms and molecules. It has developed
into a highly refined and precise method for characterization and
identification of complex molecules, as well as non-invasive imaging. Many
generations of scientists have contributed sophisticated methods of using
NMR to peer more and more deeply into molecular structure. High Temperature
Superconductivity (HTS) comes into this mature field at a relatively late
stage in its development. As such, HTS has not introduced any fundamentally
new methodology for performing NMR spectroscopy or Magnetic Resonance
Imaging (MRI). However, HTS presents the opportunity to significantly
enhance existing measurement methods by improving the signal-to-noise ratio
of the detection system. This has advantages in certain applications of NMR
spectroscopy or MRI for which the sample noise does not dominate the noise
performance of the entire measurement system.

This paper presents a simple pedagogical discussion of the impact of HTS on
NMR spectroscopy and MRI. It is written by an outsider; hence, the reader is
cautioned to consult the literature for a more complete treatment of the
subjects discussed here. A beginning student should consult Slichter's book
for the theory of NMR \cite{Schlicter}, and any one of a number of practical
introductions to NMR spectroscopy \cite{Derome,Rahman}. There is also an
excellent review paper on the use of HTS in NMR spectroscopy \cite{Hill97}.

In this paper, we begin by reviewing the fundamentals of NMR, discuss some
of the important uses of NMR in spectroscopy, and review the basic
methodology of an NMR experiment. We then discuss the requirements for
performing good NMR spectroscopy, setting the stage for the impact of HTS rf
coils on the measurement process. The conditions under which HTS rf coils
reduce system noise will be presented, followed by a discussion of the
advantages and disadvantages of using HTS rf coils. We conclude by giving an
example in which the HTS coil gives significant improvement in the
signal-to-noise ratio in real NMR spectroscopy and MRI applications.

\section{What is magnetic resonance?}

Magnetic resonance occurs under a variety of conditions. First, we require a
magnetized object, such as a ferromagnetic domain, or a collection of
paramagnetic electron spins, or a set of nuclear spins. The magnetization
can be represented by a vector ${\bf \mu }$. Next, a strong dc magnetic
field {\bf B}$_0$ is applied to the system. In the classical picture, the
magnetic field will exert a torque ${\bf \tau }=${\bf \ }${\bf \mu }\times 
{\bf B}_{{\bf 0}}$ on the magnetization or spins, causing the magnetization
vector to precess about the direction of the magnetic field at a rate
proportional to the field strength, given by $\omega _{precession}$ = $%
\gamma $B$_0$ where $\gamma $ is the gyromagnetic ratio. The gyromagnetic
ratio depends on the object carrying the magnetic moment. For electron spins
it depends on the Land\'{e} g-factor, which includes spin and angular
momentum degrees of freedom, and for nuclear spins $\gamma $ depends on the
particular nucleus studied. For example for a proton, $\gamma $/2$\pi $ =
42.5 MHz/T, while for a $^{13}$C nucleus, $\gamma $/2$\pi $ = 10.5 MHz/T.
The precessing spins create an oscillating magnetization in the plane
perpendicular to the applied magnetic field. If an rf magnetic field, polarized in
that plane, now is introduced, it can be resonantly absorbed if the rf
frequency matches the precession frequency of the spins. This resonant
absorption is magnetic resonance.

The quantum-mechanical description of magnetic resonance is as follows. The
applied magnetic field gives rise to a Zeeman splitting between ``up'' and
``down'' spin states (i.e., spin aligned along and against the magnetic
field direction). This comes about from the interaction of the magnetic
moment ${\bf \mu }$ with the applied field {\bf B}$_0$, with an energy E = -m%
$_I\gamma \hbar $B$_0$, where m$_I$ is the component of the nuclear spin in
the direction of {\bf B}$_0$, and $\hbar $ is Planck's constant divided by 2$%
\pi $. The Zeeman splitting energy between ``up'' and ``down'' states for a
spin 1/2 is given by $\Delta $E = $\gamma \hbar $B$_0$ (the proton and the $%
^{13}$C nuclei both have nuclear spin I = 1/2.) If an rf photon of energy $%
\hbar \omega _{rf}$ = $\Delta $E = $\gamma \hbar $B$_0$ is present, and the
polarization conditions are satisfied, there is resonant absorption. Both
the classical and quantum-mechanical descriptions give the same condition
for magnetic resonance.

As mentioned above, magnetic resonance occurs in many different systems:
nuclear spins (NMR), electron spins (electron spin resonance and electron
paramagnetic resonance), ferromagnetism (ferromagnetic resonance), and
antiferromagnetism (antiferromagnetic resonance). In this paper we shall
focus on NMR.

\section{What does NMR tell you?}

The precession rate of the nuclear spin is a sensitive function of the
magnetic field experienced by the nucleus. Although the applied external
field can be substantial (1 - 20 T), there are other magnetic fields at the
site of the nucleus which can contribute to the precession rate. These
fields come from electrons orbiting the nucleus, and their value will depend
on the chemical structure of the atom or molecule into which the nucleus is
incorporated. As a result, the nuclear spin precession rate will vary
slightly depending upon the chemical bonding of the electrons around that
nucleus. This variation in the precession rate, and therefore in the NMR
frequency, is called the chemical shift \cite{Derome,Rahman}. It gives
important information about the molecule and can be used as a
``fingerprint'' of the chemical structure of the molecule, because in
general, each nucleus experiences a slightly different chemical shift.

The degree of chemical shift depends on the gyromagnetic ratio of the
nucleus involved. For protons, the chemical shift range is on the order of
10 parts per million (ppm). This means that the NMR frequency will vary from
the nominal value by about 1 part in 10$^5$ due to chemical effects for a
given applied magnetic field. For $^{13}$C, the chemical shift range is
about 200 ppm.

Other information which can be obtained from NMR relates to molecular
dynamics. For instance, segments of molecules may be free to rotate, and
nuclei in these segments give information about their rate of rotation.
Other information comes from the way in which polarized nuclear spins become
de-polarized through their interactions with the environment. There are two
measurable relaxation time scales, T$_1$ and T$_2$. T$_1$ is the
spin-lattice relaxation time and T$_2$ is the spin-spin relaxation time.
Multi-dimensional NMR spectroscopy also can give information about which
nuclei are neighbors in a complicated molecule, simplifying the structural
determination and also providing unique information for molecular
identification \cite{Derome,Rahman}. All of this information contributes to
understanding the molecules under study.

\section{How is NMR done?}

An NMR experiment begins by polarizing the nuclear spins with a strong
static magnetic field. In modern NMR spectrometers and MRI systems, the
magnetic field is created by a low-T$_c$ superconducting magnet.
State-of-the-art NMR spectrometers have static magnetic fields B$_0$ which are 
typically 7.1 to 18.8 T (proton precession frequency of 300 to 800 MHz), 
MRI magnetic fields are 0.5 to 1.5 T (proton precession frequency of 21 to 63 MHz),
but low cost systems can be 0.1 T and clinical research systems can be 4.7 T.
With this field applied, the nuclear spins are set into precession at a rate
given by the field strength and the gyromagnetic ratio of the nucleus of
interest, $\omega $ = $\gamma $B$_0$. In NMR spectrometer systems the
precession frequency is typically in the range of 100 to 800 MHz. In MRI
systems, the precession frequencies are in the range of 10 to 100 MHz. Rf
magnetic fields of the correct frequency and polarization now are introduced
to manipulate the precessing spin system using rf drive coils. The rf fields
are usually pulsed. These pulses are designed to perform basic
quantum-mechanical manipulations of the spin system. The details of these
pulse sequences can be quite intricate, and the reader is referred to the
extensive and detailed literature on the subject \cite
{Schlicter,Derome,Rahman}. A simple and useful rf pulse is the $\pi $/2
pulse. This causes the spins to rotate through an angle of 90$^o$ away from
the applied magnetic field direction, causing the spins to lie in the plane
perpendicular to the applied field. This produces a maximum transverse rf
signal, which is easily detected by the pickup coils of the NMR spectroscopy
or MRI system.

If there are several types of nuclei in the sample, then each nuclear spin
precesses at a frequency depending on its gyromagnetic ratio and
environment. Very often multiple-frequency rf pulses are used to
simultaneously stimulate NMR in several different nuclear species. Once all
of the spin manipulations have been made, the system is allowed to relax in
a process called free induction decay (FID). The system uses pickup coils to
collect the FID voltage signal as a function of time. This signal is then
Fourier transformed to get the NMR spectrum of the system. The direct FID
signal can give the relaxation time scales T$_1$ and T$_2$, while the FID\
spectrum gives the chemical shift and other information.

In good spectrometers with high signal-to-noise, the NMR spectrum is
remarkably clean \cite{Derome,Rahman}. In practice, the NMR spectrum
linewidths are on the order of 1 Hz for NMR frequencies on the order of 500
MHz. This means that very small splittings, on the order of 0.01 ppm can be
resolved, making NMR spectroscopy extraordinarily information rich.

One can get an idea of what an NMR spectrometer looks like from browsing the
web sites of commercial manufacturers of spectrometers \cite
{BrukerWebSite,VarianWebSite}. The bulk of the instrument is made up of the
low-T$_c$ superconducting magnet and its associated dewar. The dewars are
remarkably efficient and may require the addition of liquid nitrogen only
once every few weeks, and liquid helium once every few months.

\section{Requirements for Good NMR Spectroscopy}

In NMR spectroscopy one often deals with small samples, sometimes smaller
than 1 microgram. Hence, the most
important consideration for an NMR spectrometer is maximizing the
signal-to-noise ratio (S/N) of the system. The most direct way to maximize
the S/N is to use the strongest static magnetic field, B$_0$, possible. It
is found that the voltage S/N ratio scales as $S/N$ $\sim \omega
_{precession}M/\sqrt{R}\sim B_0^{7/4}$ \cite{Hoult,Black93,WithersPrivate}. One
power of B$_0$ comes from the fact that the magnetization M of the
(paramagnetic) sample increases linearly with B$_0$, and a second power of B$%
_0$ comes about because the rf voltage inductively picked up by the
receive coil scales with $\omega _{precession}$, which is directly
proportional to B$_0$. However, the (normal-metal) probe resistance R
increases as B$_0^{1/2}$. In other words, as the precession frequency
increases, the skin depth in the metallic coil decreases, enhancing the
losses. Hence, the overall gain in S/N is a bit less than a power of 2.
Because of this strong scaling of S/N with static magnetic field strength,
and the fact that spectral features are more easily distinguished at high
fields (greater dispersion), there are efforts to create stronger (B$_0$ $\geq $ 20 T)
superconducting magnets for use in NMR spectrometers. However, the great
expense and engineering efforts of creating these more powerful magnets is
prohibitive, so NMR spectroscopists are looking for other ways to improve
system S/N.

HTS materials present a unique opportunity to improve S/N without great
expense or replacement of the superconducting magnet. The low resistance and
low operating temperature of HTS pickup coils significantly
increase the S/N of the detection system. The fact that the HTS pickup coils
are kept under cryogenic conditions also presents the opportunity to
introduce other cooled circuit elements such as a preamplifier, matching
network and transmit/receive switch. These considerations have made HTS
pickup coils very attractive, in lieu of an increase in static magnetic
field strength. The improvement in S/N allows smaller samples to be studied.
This in turn could reduce the cost of an NMR spectrometer because a
superconducting magnet would not have to create a homogeneous field over a
great volume.

Other considerations for good NMR spectroscopy include the need for good
sensitivity in field and frequency to distinguish closely-spaced peaks in
the FID spectrum, such as splitting coming from dipole-dipole interactions.
Good sensitivity also requires extremely good magnetic field homogeneity.
Each nucleus in the sample must see the same external magnetic field if the
small ($\sim $ 10 ppb) chemical shifts are not to be smeared. This means that
a 1:10$^9$ homogeneity of B$_0$ must be maintained over the volume of the
sample (typically a cylinder about 5 mm in diameter and 15 mm tall). The magnetic
field homogeneity is created by a series of shim coils (typically 28 or so) 
which excite magnetic fields with the symmetries of the spherical harmonics.
One also
needs good S/N and high dynamic range to see both large and small peaks in
the NMR spectrum. Finally, an improved S/N also cuts down measurement time,
a key consideration for MRI and for sophisticated multi-dimensional NMR
spectroscopy techniques.

\subsection{Overview of an NMR Spectroscopy System}

There are three parts of an NMR spectroscopy system of interest to us. The
first is the sample which contains the polarized spins producing free
induction decay. The second is the probe, which is a tuned resonant rf coil
probe for detecting the FID from the sample. The probe is a simple
inductor-capacitor (LC) resonator circuit which is tuned to the NMR signals
from specific nuclei, such as $^1$H, $^{13}$C, $^{19}$F, etc. This is the
part of the system which can be made with a large-area HTS thin film.
Finally, we have the receiver system which mixes the detected FID signal
down to lower frequencies. This part of the system can be improved with a
cryogenic preamplifier and matching network.

The probe coil has a quality factor Q, and frequency selectivity given by Q
= $\omega $L/R, where $\omega $ is the angular frequency of the NMR signal,
L is the self inductance of the coil, and R is the equivalent resistance of
the coil. Since $Q = \omega / \delta \omega$ is inversely proportional to
the 3-dB bandwidth of the resonator, $\delta \omega$, it is a measure of the
frequency selectivity of the coil.  HTS coils have the advantage of 
a significantly smaller resistance
R than their normal-metal counterparts with the same inductance. Hence, the
quality factor of the receive coils is significantly higher for HTS coils,
making them attractive for certain applications, as we shall see later.

\subsection{Noise in NMR systems}

There are three sources of noise in the NMR spectrometer detection system of
interest to us here. The first is thermal noise from conducting samples. One
can think of the sample as a resistor at finite temperature (usually room
temperature). For instance, biological samples are often in conducting
fluids. There are thermally-induced fluctuation currents in this resistor
which produce noise, called Johnson noise. Very little can be done about
this source of noise if one wants to characterize the sample (or organism)
at room temperature and in its natural state. A second source of noise is
thermal noise from resistance of the FID voltage pickup coil. Again Johnson
noise in the coil will add noise to any detected signal. A third source of
noise is noise in the receiver system, including the preamplifier, the mixer
and detector.

HTS is of benefit only when the second and/or third noise sources
dominate the system noise.\cite{Black93} In other words there is no benefit
to using HTS pickup coils in NMR or MRI systems where the sample noise
dominates the system noise.

\subsection{How Does HTS Improve Things?}

In the case where the probe and receiver dominate the noise of the NMR
system, one has the following approximate scaling relation for the
signal-to-noise ratio (S/N) of the system as the coil properties are 
varied \cite{Hill97}: S/N $\sim \sqrt{%
\frac{\eta Q}{T_p+T_A}}$, where $\eta $ is the filling factor of the coil
system (i.e. fraction of the rf energy created by the coil which is stored 
in the sample volume), T$_p$ is the effective probe temperature and T$_A$ is the receiver
noise temperature. HTS rf pickup coils have higher Q (because of their lower
surface resistance) \cite{Black95}, and they have a lower probe temperature T%
$_p$. These improvements are offset by a reduced filling factor $\eta $ for
the HTS coils. The reduced filling factor comes about because existing HTS
coils are made on flat, rigid substrates, whereas optimized normal-metal
coils are made on curved substrates which wrap around the sample. Table 1
summarizes the current state-of-the-art for the S/N improvement of HTS
pickup coils \cite{Hill97}.

\begin{table}[htb]
\caption{Signal-to-noise comparison of normal-metal and HTS rf probes}
\begin{center}
\begin{tabular}{lll}
\hline
Factor & Normal Probe & HTS Probe \\ \hline
$\eta $ & 1 & 0.2 \\ 
Q & 250 & 10,000 \\ 
T$_p$ & 300 & 75 \\ 
T$_A$ & 80 & 80 \\ \hline
S/N & 1 & 4.5 \\ \hline
\end{tabular}
\end{center}
\end{table}

In this table, the filling factor is normalized to that of the normal-metal
probe. The filling factor is about a factor of 5 worse for the HTS coil
because of present-day geometrical constraints in the deposition of the
films. However, recent progress in ion-beam-assisted deposition of HTS thin
films has created the opportunity to make HTS coils on flexible substrates.
Thus improvements in the filling factor of HTS coils should soon be possible.

In Table 1 the quality factor, Q, of HTS coils is conservatively estimated
to be about 10,000, versus about 250 for their normal-metal counterparts.
Higher quality factors in HTS coils have been demonstrated (e.g., Q\ =
70,000 at 7 T) \cite{Black93,Withers93}, and the Q appears to be limited by
flux motion \cite{Black93}. However, there are upper limits on the Q of the
coil imposed by the NMR measurement process itself - see below.

The probe temperature comparison in this table is a bit unfair. The
normal-metal probe is taken to be at room temperature, while the HTS coil
is, of course, cooled below the HTS\ transition temperature. A more fair
comparison would be between HTS coils and cooled normal-metal coils, as has
been done for the case of MRI coils \cite{Black93}. However, the amplifier
noise temperature is taken to be the same for each system. Overall, an
improvement of 4.5 in the S/N is expected simply by substituting an HTS
pickup coil for a normal-metal coil. Such an improvement in S/N has been
demonstrated in a commercial cryogenic product made by Conductus and
marketed by Varian \cite{Hill97,VarianWebSite,Wong}.  Similar improvements
have been made with the commercial cryogenic probes made by Bruker \cite{BrukerWebSite}.

\section{HTS and MRI}

The use of HTS pickup coils in magnetic resonance imaging (MRI) also is
quite promising. The most familiar example of MRI, that of whole-body
imaging, does not benefit from the introduction of HTS coils because the
noise of the sample dominates the system noise. However, low-field MRI, with
magnetic fields B$_0$ $\sim $ 0.1 T and NMR frequencies on the scale of 5 -
20 MHz, can benefit from the use of HTS rf pickup coils. These systems are
designed to look at small body parts or other small samples \cite
{Yap,Hetern,Vester,Ma99}. They also are much less expensive than full-body
MRI systems, and may find their way into other applications. However, in
order to recover the small signals generated in low-field MRI, the S/N of
the receiver system must be enhanced. HTS rf coils are a clear solution to
this challenge. Here, HTS has an additional advantage because the rf coils
are used only for receiving the FID signal; hence, issues of nonlinearity in
the material are less pressing.

In addition to standard MR imaging, there is the application of NMR
microscopy, where MRI is performed over a small volume with high resolution 
\cite{Black93,Black95,Cho}. In this case the sample (and therefore the
signal) of interest is small. System noise can have a significant
contribution from the receiver, and introduction of an HTS coil is clearly
beneficial. A simple estimate shows that the coil S/N $\sim \sqrt{Q/T_p}$,
where Q\ is the quality factor and T$_p$ is the probe temperature \cite
{Black93}. One group has demonstrated a signal-to-noise improvement of a
factor of 60 in an HTS coil at 10 K versus a Cu coil at room temperature 
\cite{Black93}. This dramatically improves the capability of the NMR
microscope.

Once again the main benefit of HTS is to improve the S/N. Improved S/N leads
to shorter acquisition time for the image because the image acquisition time
scales as (S/N)$^{-2}$ \cite{Black93}.

\section{Constraints on HTS RF Coils}

Although the above discussion makes it appear that HTS pickup coils should
be clearly superior to their normal-metal counterparts, there are serious
challenges to their introduction into NMR and MRI applications. Here, we
discuss three significant constraints imposed on HTS coils in NMR
spectrometer systems.

\subsection{Magnetic Field Perturbation}

First, the HTS coil must be introduced in such a way as to minimize the
perturbation of the static magnetic field B$_0$ \cite{Hill97,ConductusPatent}. As noted
above, the nuclei in the sample must all experience the same external
magnetic field to at least 1 part in 10$^9$. This is required to maintain
NMR spectral line widths on the order of 1 Hz, or less, for high-quality
spectroscopy. However, the introduction of a superconductor into this
magnetic field is likely to alter its distribution since the superconductor
will enter the critical state and create an inhomogeneous distribution of
magnetic flux within itself. This in turn could ruin the homogeneity of the
magnetic field in the sample.

The solution to this problem starts with use of thin-film superconductors
for the rf coils. The thickness of these films is typically less than 1 $\mu 
$m. Next, an external field B$_0$ is applied in the plane of the film,
leading to minimal disruption of the magnetic flux by the superconductor. In
addition, the superconducting coils also must be kept relatively far away
from the sample to minimize the inhomogeneity of the field at the sample.
Unfortunately, this has the effect of decreasing the filling factor $\eta $
of the coil, although this reduction is dictated to some extent by the
cryogenic constraints as well. Finally, there also is the significant
engineering challenge of keeping the superconducting rf coils aligned with
respect to the direction of the field B$_0$ as the coils are cooled from
room temperature \cite{HillPrivate}. If the coils tilt too much from the
ideal orientation, the magnetic field homogeneity at the sample position
would be lost. In addition, the HTS rf coils must not move during the pulsing
of gradient coils used in spectroscopy, when significant forces can act on 
the currents flowing in the coil \cite{HillPrivate}.  These pulses last for
milliseconds, and can excite mechanical resonances of the coil assembly.

\subsection{High Transmitter Currents}

The rf coils in NMR spectrometers often are used for both transmitting and
receiving rf pulses. This is a challenge because it requires that the coils
operate over a wide dynamic range, from high power during transmission to
low-level signal recovery during receive mode. In transmit mode, a
high-power pulse is sent to the sample. For instance, a $\pi $/2 rf pulse of
H$_{rf}$ = 1 mT for a duration of 10 $\mu $sec at a frequency of 400 MHz
requires about 10 A of current to flow through the coil \cite{Hill97}.
Superconductors have a limit to how much current they can support before
dissipation sets in. This is defined by the critical current density, J$_c$.
For this rf pulse, the film must have a critical current density J$_c$ $\geq 
$ 10$^6$ A/cm$^2$ \cite{Hill97}. Achieving such a large critical current
density in a strong magnetic field (B$_0$ $\sim $ 20 T) is not possible with
conventional low-T$_c$ superconductors. Only superconducting materials with high critical
temperatures and high upper critical fields can operate in the environment
of an NMR spectrometer. Hence, HTS thin-film superconductors are required.
In addition, the critical-current constraint means that the films must be
cooled significantly below their transition temperature to achieve J$_c$ $%
\sim $ 1 MA/cm$^2$. In practice this means that a superconductor such as YBa$%
_2$Cu$_3$O$_7$ with a T$_c$ of about 90 K must be cooled below 77 K to
operate under these conditions.

The HTS film should remain superconducting after transmitting such a pulse.
In addition, its rf properties should remain linear \cite{Black95}. This
means that the Q and probe temperature T$_p$ should remain constant during
and after rf transmit pulses. This imposes another constraint on the HTS
films. These constraints can be met through the use of epitaxial HTS thin
films.

\subsection{The Q Shouldn't Be Too High!}

In principle, the resistance of a superconductor at rf frequencies can be
made arbitrarily small at low temperatures if it has an isotropic non-zero
energy gap. This means that the Q of an rf coil made up of this
superconductor can be made arbitrarily large, since Q = $\omega $L/R.
However, larger Q is not necessarily good for an NMR spectrometer \cite
{Black95}. For an rf coil with a Q = 10,000, the probe has a bandwidth of
100 ppm at 400 MHz, limited by its 3-dB bandwidth. Recall that typical
chemical shifts of $^{13}$C are on the order of 200 ppm. Thus, a probe with
a higher Q would not be able to measure all of the chemical shifts in a $%
^{13}$C NMR spectrum. However, a larger Q may be of benefit for proton NMR,
where the chemical shifts are on the order of 10 ppm.

Another problem with a high coil Q is the long ring-down time of the rf
energy stored in the coil. The characteristic decay time for the circulating
power in the coil is given by $\tau $ = Q/$\omega _0$. For an NMR resonance
at 400 MHz with a coil Q of 10,000, the decay time $\tau $ = 4 $\mu $sec.
This means that the coil will not be able to measure the free induction
decay signal from the sample for some time after the drive pulse sequence is
ended. Higher coil Q factors would increase this ``dead time'' further. In
principle, after the pulse sequence ends, one can actively decrease the Q of
the coil to dissipate the stored energy. However, another danger is that a
high Q can decrease the characteristic time for energy transfer between the
spin system and the pickup coil. This spin damping time scale $\tau
_{damping}$ $\sim $ 1/Q, may become too short for accurate measurement of
chemical shifts or of the relaxation time scales T$_1$ and T$_2$ \cite
{Black95}.

In fact, the HTS materials do not have an isotropic energy gap. Hence they
have intrinsic losses which cannot be eliminated, limiting the ultimate Q
attainable from an HTS rf pickup coil \cite{AnlageASC}. Nevertheless, it is
believed that higher Q factors can still be achieved at rf frequencies \cite
{Black93}. To summarize, if the NMR measurements requires significant
bandwidth, the Q of the rf receive coils must be limited. Fortunately, a
high-Q rf pickup coil can always have its Q reduced in a controlled manner,
so that a variety of bandwidths can be achieved \cite{Black95,Withers93}.

\section{HTS Probes for NMR Spectroscopy}

We now are in a position to go into more detail about how HTS films are used
in an NMR spectrometer.

\subsection{Basic Probe Configuration}

A basic NMR spectrometer consists of a room-temperature sample tube
surrounded by rf coils, all of which are contained in the bore of a
superconducting magnet. There are one or more self-resonant rf coils made of
HTS films. Each coil is a simple lumped-element loop-gap LC resonator. The
loop acts as the inductance and the gap acts as a capacitor. The geometry of
the loop and the gap are adjusted lithographically to bring the resonant
frequency, $\omega _0$ = 1/$\sqrt{LC}$, equal to the NMR frequency of
interest for a given magnetic field B$_0$. A translating paddle can be used to
fine tune the resonant frequency of the rf coil after it is installed in the
probe \cite{ConductusPatent,Kotsubo}. The NMR frequency depends on the nucleus 
of interest (e.g., $^1$H, $%
^{13}$C, $^{19}$F, etc.) and the applied static magnetic field, B$_0$.
Hence, probes are designed specifically with one or two nuclei and a certain
magnetic field in mind \cite{Miller97}. Often the probes are used to excite
simultaneously one nucleus of interest, in addition to deuterium. The
deuterium signal (from a deuterated solvent) is used in a feed back circuit
to control the magnetic field B$_0$. The intricacies of rf probe design for
high-field and low-field applications are discussed in a number of papers 
\cite{Withers93,Ma99,BlackAPL,Miller,Miller96,Miller98}.

The films are typically grown on either sapphire, MgO or LaAlO$_3$ single-crystal
substrates, which have low rf losses. In addition, the substrates must have
good thermal conductivity because in use they may be cooled only
along one edge \cite{Kotsubo}. These coils typically show quality factors Q $\sim $ 10,000,
even in a field of 9.4 T, as compared to normal metal coils which have Q on
the order of several hundred to a few thousand.

\subsection{Cryogenics and Packaging}

The HTS coils must be cooled to temperatures significantly below T$_c$ to
establish a large critical current density. This must be done with a
cryogenic system which is independent of the neighboring superconducting
magnet. Closed-cycle refrigerators, such as a Gifford-McMahon cryocooler,
have been employed \cite{Hill97,BrukerWebSite,Withers93,Kotsubo}. There are several
constraints on these cryocoolers. First, vibrations of the cold head must be
minimized to maintain magnetic field homogeneity in the sample. This
requires careful vibration isolation of the compressor from the cold head.
Second, mK-level temperature control must be maintained because the rf
properties and resonant frequency of the coil are temperature
dependent. Third, the cryostat must be non-inductive, so that the rf pulses
can pass through the cryostat walls without distortion. Finally, the
cryostat also must accommodate a temperature-controlled sample holder which
is kept near room temperature. Bruker has sold probes under the name
CryoProbe$^{TM}$ which uses coils held at 15 to 30 K and a preamplifier and
transmit/receive switch at 60 to 90 K \cite{BrukerWebSite,WithersPrivate}.

\subsection{Preamplifier}

Once the signal is received by the HTS rf coils, it must be demodulated and
detected. The first step is to preamplify the signal with a low-noise
amplifier. Having cooled coils presents the opportunity to reduce the noise
temperature of the preamplifier by cooling it as well \cite{Styles,ColdPreAmps}. This
will decrease the equivalent amplifier temperature T$_A$, and increase the
system S/N. A GaAs field-effect transistor amplifier operating at 77 K has a
noise temperature of just 20 K \cite{Hill97,Black93}. The use of an all-77-K
preamplifier and cables is expected to improve the system sensitivity by
75\% \cite{Hill97}. The rf matching network also can be cooled, or even made
superconducting \cite{Withers93}. Finally, the transmit/receive switch also
can be kept at cryogenic temperature to minimize noise and losses \cite
{WithersPrivate}.

\section{Results with HTS Coils}

\subsection{NMR Spectroscopy}

There are several very convincing demonstrations that HTS provides
significant S/N improvements in NMR spectroscopy and MRI applications.
Bruker Instruments has demonstrated a factor of 4 improvement in S/N in the
their 500-MHz and 600-MHz $^1$H and $^{19}$F cryogenic probes, as well
as $^{13}$C probes \cite
{BrukerWebSite}. These improvements are along the lines of those estimated
above. This translates to the ability to measure samples with one fourth the
mass, or to measure in one sixteenth the time required with a conventional
NMR probe. The measured NMR linewidths are almost as good as those measured
by the conventional probe. These cryogenic probes are commercially available.

NMR spectroscopists use many sophisticated techniques to learn about the
interactions of nuclear spins in complex molecules. One of these techniques
is called Total Correlation Spectroscopy (TOCSY). This is basically a way of
examining how spin polarization diffuses through a molecule due to
nucleus-nucleus interactions. Hill has demonstrated that a Conductus HTS probe can
significantly increase the S/N of a TOCSY spectrum - see Fig. 5 in Hill's
paper \cite{Hill97}.

Another spectroscopy method, called Heteronuclear Multiple Quantum Coherence
(HMQC), excites two different nuclei simultaneously (e.g., $^1$H and $^{13}$%
C, this is an example of two-dimensional NMR) and establishes which of these
nuclei are directly bonded to each other. This kind of information can be
used as a fingerprint to uniquely identify a complex molecule. However,
since the $^{13}$C isotope is so rare (1.1\% natural abundance), it takes 
considerable time (on the
order of 12 hours) to acquire enough data to establish the HMQC spectrum.
This is a situation for which HTS rf coils provide a clear advantage. With
the improved S/N brought by HTS, and good long-term stability, the HMQC
spectrum can be acquired in much less time. Bruker has a very clear dramatic
demonstration of this improvement on their Web site for the cases of HSQC
and HMBC two-dimensional NMR \cite{BrukerWebSite}.

With the significant improvements in S/N brought by HTS coils, it may be
possible to develop altogether new types of multi-dimensional NMR
spectroscopy which would be impractical using conventional rf coils. Other
companies thought to be investigating cryogenic and/or HTS rf coils for NMR
spectroscopy include Dupont, IGC and Nalorac.

\subsection{MRI}

The improved signal-to-noise ratio of HTS pickup coils can significantly
decrease image acquisition time in MRI systems. Penn, {\it et al}. have
demonstrated a factor of 3 improvement in S/N using an HTS coil versus a
silver coil, for both at 77 K in a 0.15-T low-field MRI system \cite{Penn}.
A factor of 10 improvement in S/N was demonstrated by Black, Early and
Johnson in an HTS pickup coil versus a room-temperature copper coil in an
NMR microscope with 30 $\mu $m resolution \cite{Black95}. A factor of 5
improvement of S/N in MR imaging of a rat spine was found by Wosik, {\it et
al. }\cite{Wosik}

MRI of sodium has been useful for examining cell integrity and has
advantages over proton MRI for detecting certain kinds of brain disorders.
However, sodium MRI suffers from lower S/N, making HTS pickup coils
attractive for this application \cite{Miller}. A very interesting market
analysis of the impact of HTS on MRI has recently been published \cite{Smith}%
.

\section{Conclusions}

We have given a simple pedagogical overview of the impact of HTS on NMR
spectroscopy and magnetic resonance imaging. Although HTS does not introduce
any radically new measurement process to this field, it does make
significant improvements to the signal-to-noise of the pickup system. This
comes at a time when other improvements to signal-to-noise are becoming
prohibitively expensive to pursue. Commercial products which employ
cryogenic technology in the rf pickup coils are now available and have
clearly established their superiority over normal-metal rf probes. Further
improvements can be made to the HTS coils, most notably with films grown on
flexible substrates for improved sample filling factor.

The author has benefited from fruitful discussions with Richard Withers of
Bruker Instruments, Howard Hill of Varian, and Doran Smith of the Army
Research Laboratory. Dr. Withers is also acknowledged for a critical reading
of the manuscript.  The author acknowledges the support of the NATO ASI on
Microwave Superconductivity, as well as the Maryland Center for
Superconductivity Research.

% References

\end{document}